\documentclass[pdf,aps,pra,twocolumn,notitlepage,floatfix, superscriptaddress, nofootinbib, letterpaper,longbibliography]{revtex4}

 \usepackage[pdftex, breaklinks, plainpages=false,colorlinks=true,citecolor=blue,linkcolor=blue,urlcolor=blue,filecolor=green,bookmarksopen=true]{hyperref}
  \usepackage{amsmath, amstext, amssymb, amsfonts,amsmath, mathtools}
  \usepackage{natbib, url, textcase, bm} 
  \usepackage{microtype, graphicx, algorithm, algpseudocode}
  \usepackage{subfigure}
  \usepackage[english]{babel}
  \usepackage{exscale}
  \usepackage[dvipsnames]{xcolor}

\begin{document}

\title{An Application of Quantum Annealing Computing to Seismic Inversion} 
 \author{Alexandre M. Souza}\email{amsouza@cbpf.br}
   \affiliation{Centro Brasileiro de Pesquisas F\'isicas -- CBPF, Rua Dr.\ Xavier Sigaud 150, 22290-180 Rio de Janeiro, Brazil }  
\author{ Eldues O. Martins}
 \affiliation{Centro de Pesquisa e Desenvolvimento Leopoldo Am\'erico Miguez de Mello -- CENPES/PETROBRAS, Av.\ Hor\'acio Macedo 950, 21941-915 Rio de Janeiro, Brazil}

\author{ Itzhak Roditi}
  \affiliation{Centro Brasileiro de Pesquisas F\'isicas -- CBPF, Rua Dr.\ Xavier Sigaud 150, 22290-180 Rio de Janeiro, Brazil }
   \affiliation{Institute for Theoretical Physics, ETH Zurich 8093, Switzerland }
\author{ Nahum S\'a}
  \affiliation{Centro Brasileiro de Pesquisas F\'isicas -- CBPF, Rua Dr.\ Xavier Sigaud 150, 22290-180 Rio de Janeiro, Brazil }
\author{ Roberto S. Sarthour}
\affiliation{Centro Brasileiro de Pesquisas F\'isicas -- CBPF, Rua Dr.\ Xavier Sigaud 150, 22290-180 Rio de Janeiro, Brazil }
\author{ Ivan S. Oliveira }  
\affiliation{Centro Brasileiro de Pesquisas F\'isicas -- CBPF, Rua Dr.\ Xavier Sigaud 150, 22290-180 Rio de Janeiro, Brazil }
%



\begin{abstract}
Quantum computing, along with quantum metrology and quantum communication, are disruptive
technologies that promise, in the near future, to impact different sectors of academic research and industry. Among the computational challenges with great interest in science and industry are the inversion problems. These kinds of numerical procedures can be described as the process of determining the cause of an event from measurements of its effects. In this paper, we apply a recursive quantum algorithm to a D-Wave quantum annealer to solve a small scale seismic inversions problem. We compare the obtained results from the quantum computer to those derived from a classical algorithm. The accuracy achieved by the quantum computer is at least as good as that of the classical computer.


\end{abstract}
\maketitle

-----------------------------------------------------------------------
-----------------------------------------------------------------------
 \section{Introduction}
Seismic geophysics relies heavily on subsurface modeling based on the numerical analysis of data collected in the field. The computational processing of a large amount of data generated in a typical seismic experiment  can take an equally large amount of time before a consistent subsurface model is produced.
Electromagnetic reservoir data, like CSEM (Controlled Source Electromagnetic), petrophysical techniques, such as electrical resistivity and magnetic resonance on multi-wells, and engineering optimization problems like reservoir flux simulators, well field design and oil production maximization also need a strong computational apparatus for analysis.

On the other hand, in the past decade, there has been much progress in the development of  quantum computers: machines exploiting the laws of quantum mechanics to solve hard computational problems faster than conventional computers.
A concrete example of such progress is the so-called quantum supremacy, that has been recently demonstrated using specific purpose quantum computers \cite{google,boson,china}. The Geoscience field  and related industries, such as the hydrocarbon industry,  are strong candidates to benefit from those advances brought by quantum computing.

Currently, different quantum technologies and computational models are being advanced.
Giant companies like IBM, Google, and Intel are developing quantum computers based on superconducting technologies \cite{oliver}.
Other companies are also putting considerable effort into building a fully functional quantum computer based on Josephson junctions, such as the North American Rigetti, whereas, the also American IonQ and the Austrian AQT  are working on computers based on trapped ions \cite{sage}. The Canadian company D-Wave, leader in the computational model known as quantum annealing \cite{Job}, is already trading quantum machines, and the also Canadian Xanadu is providing cloud access to their  photonic quantum computer \cite{bourassa,brod}.  Recent reviews comparing superconducting and trapped-ion technologies and different cloud-based platforms can be found in references  \cite{linke} and \cite{devitt}, respectively.

 In the field of Geoscience, recent works have used quantum annealers to hydrology inversion problems \cite{hydrology1,hydrology2}. In those works, it was shown that, although the size of the problem that can be solved on a third-generation  D-Wave quantum annealer is considered small for modern computers, they are larger than the problems solved with similar methodology with Intel's  third and fourth generation chips.  It is also important to mention that optimization techniques are widely used in seismic inversions, but usually classical algorithms  get stuck in local minima.  Previous works \cite{sen,wei,greer} have indicated that quantum annealing can be advantageous to solve seismic problems. However, the potential applications of quantum computing in Geoscience has so far been largely unexploited in the specialized literature.  
 
 In this work, we present a formulation of a seismic problem as a binary optimization, and one small scale subsurface seismic problem is solved using the D-Wave quantum annealer available in the Amazon Braket service.  We have evaluated the performance of the quantum computer comparing the results obtained to those derived from a classical computer. Our analysis was focused on the accuracy. It was found that the accuracy achieved by the quantum computer is at least as good as that of the classical computer for the problem we have studied.  

This paper is organized as follows. In Sec. \ref{annealing} we introduce the basic idea of quantum annealing. In Sections \ref{seismic_qubo} and \ref{methods},  we present the formulation of a seismic inversion as a binary optimization problem and the methods used, respectively.  In Sec \ref{result},  the results obtained in the D-Wave quantum annealer are shown. In the last section, we draw the conclusions.
 
 \section{Quantum Annealing} \label{annealing}
In the literature, there are many different quantum computational models developed. Currently, three main models of quantum computing are being considered: the logic gate model, or circuit model, boson sampling and the adiabatic model. The gate model is an universal quantum computation model that is  performed programming a step-by-step instruction build from basic building blocks, known as quantum gates, similar to the classical circuit model \cite{nielsen}. This model is exploited by companies such as IBM, Intel, Rigetti, IonQ, AQT, and Google. Boson sampling computers consists in sampling from the output distribution of bosons in a linear interferometer \cite{bsampling,brod}. There is strong evidence that such an experiment is hard to be simulated in classical computers, but it is efficiently solved by special purpose photonic chips. The adiabatic computation model \cite{Farhi,lidarRMP} is a model in which the computational problem is mapped into a Hamiltonian, in such a way that the solution of the computational problem is encoded in the ground state of the quantum system represented by the Hamiltonian  $H_{final}$.  The computation is performed starting from the ground state of a known Hamiltonian $H_{initial}$. The Hamiltonian is slowly modified towards to the target Hamiltonian $H_{final}$. During the process, the total Hamiltonian of the system is given by  
 \begin{equation}
 H(p) = (1-p)H_{initial} +p H_{final}  \label{eqadiabatic},
 \end{equation}
 where $p \in [0,1]$. According to the adiabatic theorem, if the evolution is performed adiabatically, the quantum state of the system remains in the instantaneous ground state throughout the entire process. 
 
The adiabatic model is equivalent to the gate model \cite{aharonov}, i.e., any problem that can be computed in the gate model will also be computable in the adiabatic model.  This statement is valid for certain types of  $k-$local Hamiltonians \cite{aharonov}. For example, considering a system of $n$ qubits, we can perform universal adiabatic computation by choosing \cite{biamonte}
\begin{eqnarray}
H_{initial} &=& \sum_i \delta_i \sigma_i^x, \\
H_{final} &=& \sum_i H_i \sigma_i^z +  \sum_{j>i}  J_{i,j} \sigma_i^z \sigma_j^z + \sum_{j>i}  K_{i,j} \sigma_i^x \sigma_j^x,
\end{eqnarray}
 where $\sigma_i^k$, with $k=x,y,z$, is one of the Pauli matrices of the $i^{th}$ qubit, $\delta_i$ and $H_i$ are local transverse and longitudinal fields, respectively,  and $J_{i,j}$ and $K_{i,j}$  are coupling constants. 
   
The adiabatic model can be viewed as a special case of the quantum annealing computing. In quantum annealers,  the Hamiltonian change is not adiabatic. Therefore, quantum annealing is a heuristic type of computation. The D-Wave computer is a quantum annealer that uses Ising chains
\begin{eqnarray}
H_{final} &=& \sum_i H_i \sigma_i^z +  \sum_{j>i}  J_{i,j} \sigma_i^z \sigma_j^z.  \label{ising}
\end{eqnarray}
The quantum annealing performed with Ising chains is unlikely to implement universal quantum computation \cite{preskill}. Therefore, D-Wave annealers are more restrictive than the universal adiabatic model. Quantum annealing with Ising chains can be applied to a class of computational problems known as NP-hard problems \cite{barahona}. It is believed that quantum annealing will be able to find better approximate solutions or find such approximate solutions faster then classical computers \cite{preskill}.  Here, it is also important to mention that the advantage of quantum annealers over classical methods is still under debate \cite{ronnow,mandra,Katzgraber}. Recent works have proposed that, in some cases,  quantum annealing is advantageous over classical computing \cite{albash, denchev,Li2, baldassi}, on the other hand,  no advantage was reported in references \cite{ronnow,hen}.   The origin of the possible speedup is also under debate. Quantum tunneling is often claimed to be the key mechanism underlying possible speedups of quantum annealing. However, recent work has found  numerical evidence that  quantum tunneling processes can be efficiently simulated by Monte Carlo methods  \cite{brady} . There is also evidence to suggest that it is unlikely to achieve exponential speedups over classical computing solely by the use of quantum tunneling \cite{crosson}.   The role of the temperature in the performance of quantum annealing has been  also studied in \cite{jiang}.

To solve a problem in the D-Wave quantum annealer \cite{Job}, it is necessary first to express the problem to be solved as an Ising problem or as a quadratic unconstrained binary optimization (QUBO), which is equivalent to Ising but defined on the binary values $0$ and $1$, whereas the Ising problem is defined on the binary values  $\pm 1/2$. The QUBO problem can be written as the minimization of the quadratic function 
\begin{equation}
f(\mathbf{q}) =\mathbf{q}^\mathsf{T} \mathbf{Q q},   
\label{fun}
\end{equation}
where $\mathbf{q} \in \{0,1\}^n$,  $\mathbf{Q}$ is a $n \times n$ upper (lower) triangular matrix and the vector  $\mathbf{q}^\mathsf{T} = (q_1,q_2, \cdots, q_n)$ contains $n$ binary variables. QUBO problems are commonly used in machine learning and many important computational problems can be translated to a QUBO formulation as well  \cite{Lucas}.  Examples of problems that have been  addressed with a D-Wave quantum annealer are: the classification of human cancer types \cite{Li}, traffic optimization  \cite{traffic2}, transcription factor DNA binding \cite{Li2},  metamaterial designing \cite{kitai} and Higgs boson data analysis  \cite{mott}.

Recently, there has been a growing interest in quantum algorithms for systems of linear equations, $\mathbf{Ax=b}$, where $\mathbf{A}$  is a $n \times n$ matrix and $\mathbf{b}$ is a unit vector. Such algorithms may find applications in different research areas, including Geoscience. In the quantum gate model,  the quantum version of such problem is called Quantum Linear Systems Problem (QLSP) \cite{duan,wossing2}, and it is defined as the problem of preparing the state 
\begin{equation}
|\psi\rangle = \frac{\sum_i^n x_i |i \rangle}{\sqrt{\sum_i^n |x_i|^2}} \label{psix},
\end{equation}
where  $\mathbf{x} = (x_1, x_2, \cdots)^\mathsf{T}$  is the solution of $\mathbf{Ax=b}$. 

In 2008, Harrow, Hassidim, and Lloyd (HHL) proposed a quantum algorithm for the QLSP problem \cite{hhl}. Given some assumptions \cite{aaronson}, the run time of HHL is $O (k^2 s^2 \log(n)/\epsilon)$ where $k$ is the condition number of the matrix $\mathbf{A}$, defined as the absolute value of the ratio between the largest and smallest eigenvalues of  $\mathbf{A}$, $s$ is the sparsity of $\mathbf{A}$, defined as the number of nonzero entries per row, and $\epsilon$ is the desired precision. 

After the initial HHL proposal, several improvements were achieved: the condition number dependence was reduced from $k^2$ to $k \log^3(k)$ \cite{ambainis}, the error dependency was reduced from $1/\epsilon$ to a polynomial function in $log(1/\epsilon)$ \cite{childs}, and a sparsity-independent runtime scaling was achieved in \cite{wossing}.  The QLSP problem can also be solved using iterative quantum solvers in runtime $O( k^2 \log(k) /log(1/\epsilon))$  \cite{shao} and with runtimes $O( k \log(k) /\epsilon)$ and $O( k^2 \log(k) /\epsilon)$ using the evolution randomization method, a simple variant of adiabatic quantum computing where the parameter $p$ in (\ref{eqadiabatic}) varies discretely, rather than continuously \cite{subasi}.  The best general-purpose classical conjugate gradient  algorithm  to solve $\mathbf{Ax=b}$ has the runtime $O(nks\log(1/\epsilon))$. Here, we must emphasize a fundamental difference between classical and quantum algorithms. While the conjugate gradient returns the solution vector $\mathbf{x}$,  quantum algorithms return a quantum state, equation (\ref{psix}),  that approximately contains all the components $x_i$ of the solution vector $\mathbf{x}$.  It is possible to obtain any specific
entry  $x_i$  by measuring the output state  (\ref{psix}), but  in general, it will require repeating the algorithm  many  times, which would  kill the exponential speedup.  Still,  quantum algorithms can be used as subroutines in different applications \cite{aaronson,duan}.

In quantum annealers, the problems of solving a system of linear equations and a system of polynomial equations were previously studied in \cite{rogers,borle,malley} and \cite{chang}, respectively. Unlike the previously mentioned quantum algorithms,  quantum annealing solves $\mathbf{Ax=b}$ completely, i.e.,  it returns the vector $\mathbf{x} $.  To compare the performances of a quantum annealer and a classical computer, we must take into account the cost to prepare the problem, i.e., the procedure to map $\mathbf{Ax=b}$ into a QUBO or Ising problem,  the cost to perform the annealing, and also the cost to post-process the results.  The performance of quantum annealers to solve linear systems was studied in detail in reference \cite{borle}. It was shown that quantum annealers might be competitive if there exists a post-processing method that is polynomial in the size of the Matrix  $\mathbf{A}$  with a degree less then 3.

\section{Seismic Inversion written as a QUBO problem} \label{seismic_qubo}
\begin{figure}
\centering
   \includegraphics[width=1.0\linewidth]{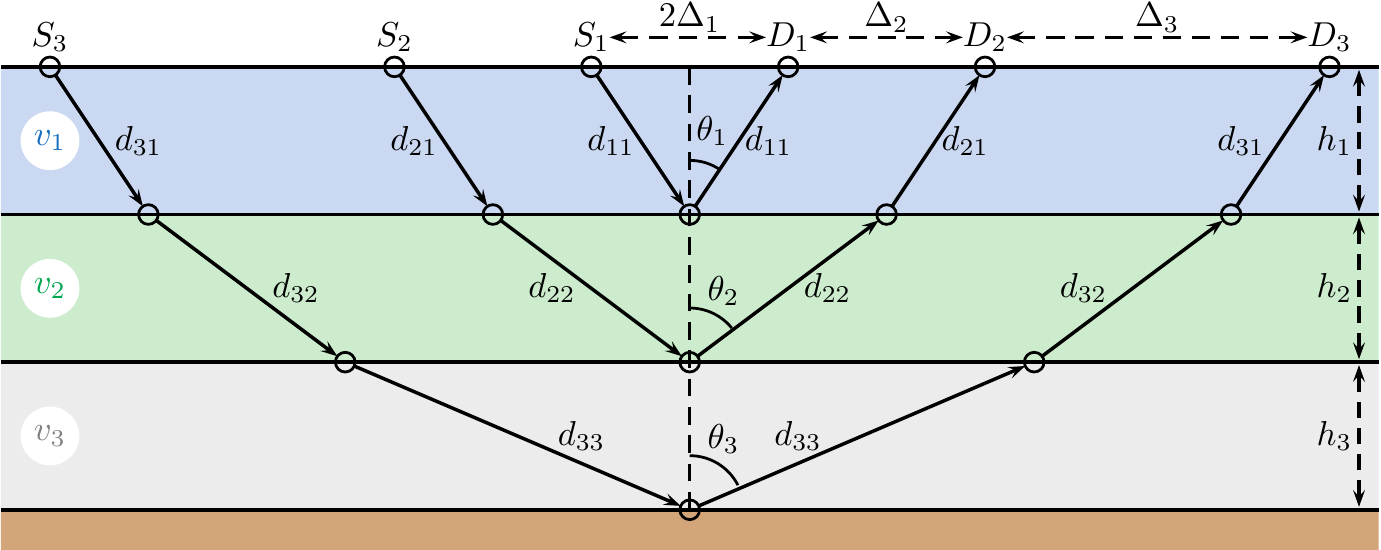}
\caption{Schematics of seismic data acquisition.  Sound traces created in the $ i^{th}$ source   $S_i$  are detected at $D_i$ after the time of travel $t_i$. This figure depicts a simplified scheme with only $m = 3$ layers, but we have considered up to $m = 46$.}
\label{sismica}
\end{figure}

We considered the propagation of sound waves in a multi-layered medium, as shown in Figure \ref{sismica}. Multiple sources produce sound waves that can be reflected in the interface of each layer. Assuming that the wave propagation can be modeled as narrow beams or rays,  the sound trace originated in the $i^{th}$ source reaches the $ i^{th}$ detector after the time interval 
\begin{equation}
t_i =2  \sum_{j=1}^i  \frac{d_{ij}}{v_j  \label{time}}, 
\end{equation}
where $d_{ij}$ and $v_j$ are the distance traveled by the sound waves and the sound speed in the $j^{th}$ layer, respectively. If we consider the thickness of each layer as $h_j$ and the distance between two consecutive sources (detectors) as $\Delta_i$, we can write $d_{ij} = h_j/\cos \theta_j$ where $\theta_j = \arctan(\Delta_j/h_j)$.

 The layered model described above is commonly used in seismic explorations, either offshore or onshore \cite{geo}. In seismic experiments  the goal is to determine the velocities $\{v_j\}$ from the time intervals $\{t_i\}$,  by solving the system 
\begin{equation}
 \mathbf{Ms=t},
\label{system}
\end{equation}
 where $\mathbf{t} ^\mathsf{T}= (t_1,t_2, \cdots, t_m)$, $\mathbf{s}^{\mathsf{T}}  = (1/v_1,1/v_2 \cdots, 1/v_m)$ is the slowness vector,  $\mathbf{M}$ is a $m \times m$  lower triangular matrix with nonzero elements given by $M_{.,j}=2h_j/\cos\theta_j$, and $m$ is the number of layers.

In order to use a quantum annealer to solve the above seismic problem it is necessary to translate the problem into a QUBO formulation. To proceed, first, we rewrite the system (\ref{system})  as a minimization problem with the objective function
\begin{equation}
f(\mathbf{s}) = ||\mathbf{M}\mathbf{s} - \mathbf{t}||^2, 
\label{fun}
\end{equation}
where $\mathbf{s} \in \mathbb{R}^m$.  Next we write the slowness vector as  $\mathbf{s}=\mathbf{s_0}+L(\mathbf{x-I})$,  where $L$ defines the bounding limits of $\mathbf{s}$,  $0 \leq x_i < 2$  $\forall$  $x_i$, $\mathbf{I}^\mathsf{T}=(1,1, \cdots)$ and  the vector $\mathbf{s_0}$ is an initial guess for $\mathbf{s}$.   The objective  function $f(\mathbf{s})$  is rewritten as  
\begin{equation}
f(\mathbf{x}) =  ||\mathbf{M}\mathbf{x} - \mathbf{b}||^2,
\end{equation}
where $\mathbf{x} \in \mathbb{R}^m$ and  $\mathbf{b} = (\mathbf{t}  + L \mathbf {M I} - \mathbf{M s_0 })/L$.   The matrix $\mathbf{M}$ and the vector $\mathbf{b}$ are parameters of the objective function while the vector $\mathbf{x}$ must be converted into a binary format. Here we discretized each  $x_i$  variable with the R-bit  approximation 

\begin{equation}
x_i =\sum_{r=0}^{R-1} q_{i,r}2^{-r}. \label{rbit}
\end{equation}
 To formulate our QUBO problem, we construct a new binary vector $\mathbf{q}$ and a new real matrix $\mathbf{A}$ in order to form the binary system of equations  $\mathbf{Aq = b}$. It is straightforward to reformulate this system as a binary optimization problem \cite{rogers,borle,malley}, whose solution vector, $\mathbf{q^*}$, is given by
\begin{eqnarray}
  \mathbf{q^*} &=& \arg  \min_{\mathbf{q} \in \{0,1\}^{R \times m}}||\mathbf{A}\mathbf{q} - \mathbf{b}||^2 \\ 
                        &=& \arg \min_{\mathbf{q} \in \{0,1\}^{R \times m} } \mathbf{q}^\mathsf{T} \mathbf{Q q} + C,
\label{funBin}
\end{eqnarray}
where 
\begin{eqnarray}
 Q_{ii} &=& \sum_{k=1}^{m}{A_{ki}A_{ki}-2A_{ki}b_k}, \\
 Q_{ij} &=& \sum_{k=1}^{m}{2A_{ki}A_{kj}},  \text{       for  }   i < j,  
\end{eqnarray}
and  $C =\sum_{k=1}^{m} b_k^2$ is an additive constant that does not change the ground state.
 
The precision of the solution depends on how many binary digits are used to represent the real variables of the problem,  a solution with good precision would consume a large number of qubits of the quantum hardware. Here we have used a recursive approach similar to what was used  in \cite{rogers} to improve the precision  of floating-point division. Our recursive approach is described in the algorithm (\ref{recur}),  using such an approach we could  improve the solution of a system of linear equations with 46 real variables, using just a few qubits to represent each variable.  Next, we will show in our example that using $R =3$, in equation (\ref{rbit}), and carrying $20$ iterations is sufficient to reach a good solution.

\section{Methods} \label{methods}

We have performed the quantum computation with the D-Wave computer provided by the Amazon Braket service. Currently, there are two versions of quantum annealer available in Amazon Braket. The first is the D-wave 2000Q version, this computer contains  2041 working qubits. The connections among the qubits are represented  by a graph called Chimera \cite{bunyk}.  In this topology, each qubit is coupled to no more than $6$ other qubits. We can call Chimera as a graph of degree $6$. The second version is the D-Wave Advantage system, it is a more advanced computer with 5436 working qubits disposed in a Pegasus graph with degree 15 \cite{pegasus}. A QUBO problem can also is represented by a graph, where each vertex of the graph corresponds to a binary variable $q_i$.  When the QUBO problem is represented by a graph with degree greater then 6, for Chimera, or 15, for Pegasus, it is necessary to embed the QUBO graph onto the chip topology. The present seismic problem, for example,  is represented by a full graph, where each vertex is connected to all other vertices. In this work, the embeddings were obtained using  a heuristic algorithm \cite{cai} provided by D-Wave.
 
Errors during the computation are important issues to be considered. In quantum annealing, the computational problem is encoded into the ground state of the Ising Hamiltonian (\ref{ising}), the gap between the ground state and the excited states is a key property of the system. If the gap is too small, thermal excitations and non-adiabatic transitions can induce transitions, as a consequence, the computer will output an excited state, which can be viewed as a computational error \cite{pearson}. In addition, the wrong implementation of the Hamiltonian (\ref{ising}) may result in the wrong ground state \cite{pearson}. We have noticed that the ground state was achieved with high probability. Therefore, unwanted transitions due to thermal fluctuations or non-adiabatic evolution do not represent an important issue in our case. The annealing time used was 20 $\mu s$, the minimal and default value of the D-Wave machine. To post-process the output we use default D-Wave' routines.

In our implementation, analog errors are the most important, i.e., inaccurate implementations of the parameters $H_i$ and $J_{i,j}$ described in the equation (\ref{ising}). To reduce the impact of analog errors, we have used $10$ spin-reversal transforms, also known as gauge transformations \cite{qpu,hristo}. This type of transformation is based on the fact that the structure of the Ising problem is not affected when the following transformations are applied: $H_i \rightarrow g_i H_i$ and $J_{i,j} \rightarrow g_ig_j J_{i,j}$, where $g_i = \pm 1$. The original and transformed problems have identical energies. However, the sample statistics are affected by the spin-reversal transform because the quantum hardware is a physical object subject to errors.

\section{Results} \label{result}

\begin{figure}
\centering
  \includegraphics[width=1.0\linewidth]{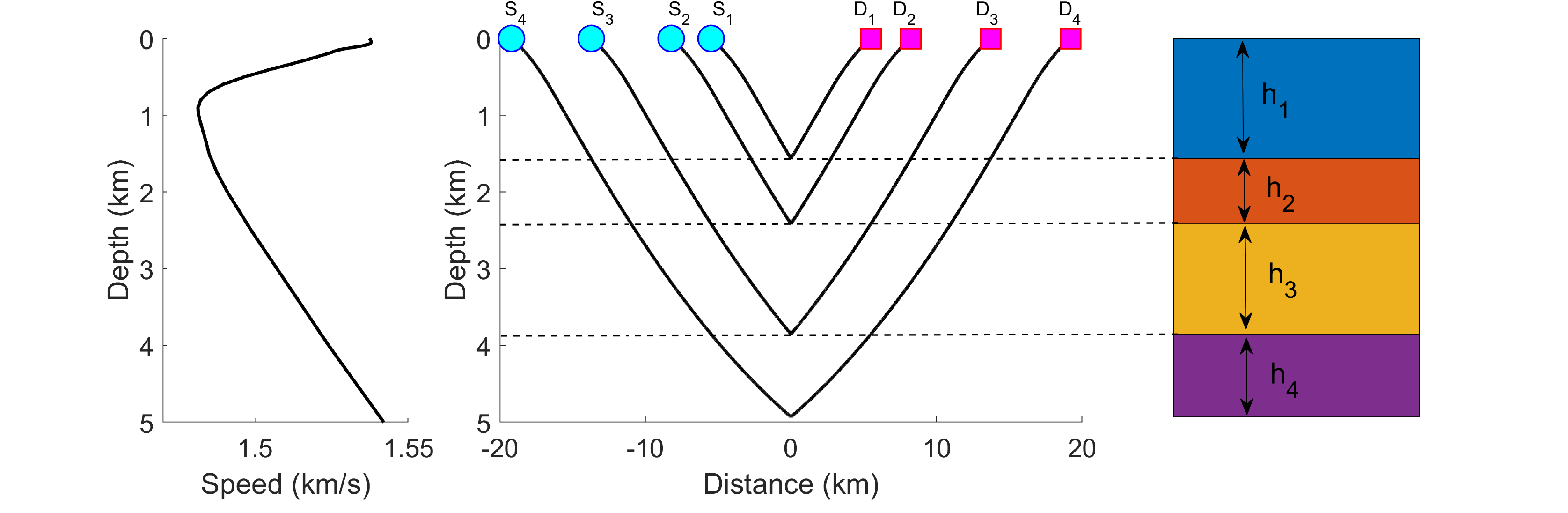}
\caption{  Acoustic sound traces. In the left panel, we show the sound speed profile of the Philippine sea \cite{hodges}. In the middle panel, we show the results of  numerical simulations of sound traces with incident angle $\theta_0 = 80^o$ and, in the right panel, we show the seismic model used. The thickness of each layer of the model matches the position where the corresponding sound trace is reflected. For clarity, we only show the simulations involving only 4  sources (receptors).}
\label{ocean}
\end{figure}

 We have applied the above formulation to solve a small scale underwater seismic inversion problem.  Artificial data were generated by simulating sound traces in  ocean. We have used the sound speed profile of the Philippine Sea, available to public \cite{hodges}, as shown in Figure \ref{ocean}.  From the simulation, we obtained the travel times between the sources and detectors. The seismic inversion model was constructed using 46 layers while the real variables were digitalized with $R =3$ bits. This model yields 138 binary variables,  for the present type of problem, it is the maximum number of variables that we can embed in the working graph of the D-Wave Advantage system available in Amazon.

\begin{figure}
\centering
 \includegraphics[width=1.0\linewidth]{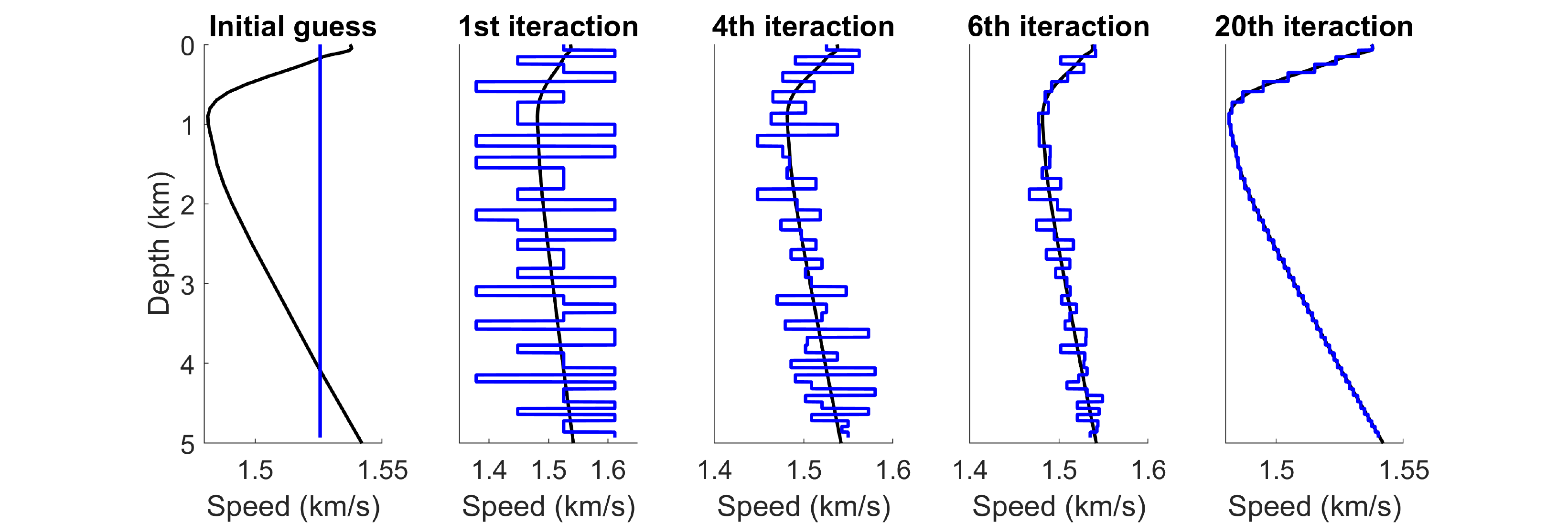}
\caption{Seismic inversion obtained in the quantum annealer for different iterations. The black curve is the original sound speed profile and the blue curve is the result of the inversion obtained in the quantum annealer. }
\label{iter}
\end{figure}

To perform the seismic inversion we have considered that all layers in the model match the position where the sound wave is reflected,  as shown in Figure \ref{ocean}.  The solution is presented in Figure \ref{iter}, as can be seen, good results are obtained with 20 iteractions.   We have also performed the inversion with a classical computer running the forward substitution algorithm to invert the lower triangular matrix in equation (\ref{system}). When classical and quantum inversions are compared, we found that the relative error between them  was $\approx 10^{-4}$, as shown in figure \ref{iter}. This result shows that the quantum computer using the recursive approach to solve a system of linear equations has enough control to find solutions with good precision.  

The results could also be compared to the benchmark provided by the condition number of the numerical problem at hand. When performing a numerical inversion procedure on the lower triangular matrix $\mathbf{M}$ to recover the solution $\mathbf{s}$ in Equation (\ref{system}), a straightforward estimation of the lower bound of the relative error $\epsilon_s$ in $\mathbf{s}$ arising from the relative error $\epsilon_t$ in the righthand side vector $\mathbf{t}$ due to the numerical conditioning of $\mathbf{M}$ is given by  \cite{kincaid}

\begin{equation}
\label{relative_errors}
\epsilon_s = \kappa_\infty \epsilon_t,
\end{equation}
where  $\kappa_\infty$  is the condition number given by
\begin{equation}
\kappa_\infty \geq \frac{\max_j (h_{j}/\cos \theta_j)}{\min_j(h_{j}/\cos \theta_j)}.
\end{equation}

Therefore, the condition number relates the error associated with $\mathbf{t}$, $\epsilon_s$, and the error of the solution $\mathbf{s}$, $\epsilon_s$. In this particular application of seismic inversion, where distances are generally in the scale of $10^{3}m$, it is reasonable to assume that the figures are accurate up to the order of $\approx 10m$, with the next order of magnitude ($1m$) giving the scale of the error $\epsilon_t$. Given that the error $\epsilon_t$  can be estimated as  $\approx 10^{-3}$ and the condition number is $\kappa_\infty \approx 1$, then we can estimate $\epsilon_s \approx 10^{-3}$ from (\ref{relative_errors} ). This shows that the error of the quantum approach comes from the inversion problem by itself and is of the same order as the classical approach.

\section{Conclusions} \label{conclusion}

Quantum computing represents a fundamentally different paradigm, an entirely different way to perform
calculations. The Geoscience field and related industries are strong candidates to benefit from it. However, the performance  of Geoscience inversion problems  on current available quantum computers has so far been largely unexploited.  

In this paper, we have solved a seismic inversions in a D-Wave quantum annealer. The seismic problem was written as a system of linear equations and then translated into a QUBO formulation. The results presented here indicate that the current available quantum annealers can solve a seismic inversion at a relatively small size with nearly the same accuracy as a classical computer. The proof-of-principle computations performed here show some promise for the use of quantum annealing in Geosciences. 

The practical use of quantum annealing in Geosciences will require the ability to solve large problems.  To address this issue, decomposer tools have been proposed to divide a large problem into small subproblems which can be solved individually by the quantum hardware  \cite{qbsolv,okada,nishimura}. Using such approach, it is possible to solve a large-sized problem using just a limited number of available qubits.  Another interesting approach is the reverse annealing.  Within this method, one starts from known local solutions which can be obtained in a classical computer.  The annealing is performed backward from the known classical state to a state of quantum superposition, then proceeding forward it is possible to reach a new classical state that is a better solution than the initial one.  Recently,  it has been shown that it is possible to refine local solutions with recursive applications of reverse annealing \cite{passarelli,yamashiro,venturelli,arai}. We believe that the development of hybrid quantum-classical methods, such as mentioned above, will be essential to solve complex seismic problems on quantum annealers in the near future.

Finally, we should mention that the problem solved here is well-conditioned, however, often Geoscience problems are ill-conditioned. An interesting question is whether quantum computers can solve ill-conditioned problems efficiently. In reference \cite{clader}, it was theoretically proposed that preconditioning methods can expand the number of  linear systems  problems that can achieve exponential speedup over classical computers. In future works, we plan to study the performance of the quantum annealers to solve ill-conditioned problems.  Another attractive prospect for future work is the implementation of Geoscience problems in gate model computers, based either on superconducting or trapped ions technologies.

\section*{Conflict of Interest Statement}
The authors declare that the research was conducted in the absence of any commercial or financial relationships that could be construed as a potential conflict of interest.

\section*{Author Contributions}
A. M. Souza ran the quantum computer, analyzed the results, co-wrote, and reviewed the manuscript. E. O. Martins initiated the study, co-wrote, and reviewed the manuscript.  N. Sá,  I. Roditi , R. Sarthour  and I. S. Oliveira co-wrote and reviewed the manuscript.

\section*{Funding}
This work was supported by the Brazilian National Institute of Science and Technology for Quantum Information (INCT-IQ) Grant No. 465469/2014-0, the Coordenação de Aperfeiçoamento de Pessoal de Nível Superior - Brasil (CAPES) - Finance Code 001, Conselho Nacional de Desenvolvimento Científico e Tecnológico (CNPq) and  PETROBRAS: Projects 2017/00486-1, 2018/00233-9 and 2019/00062-2. A. M. S. acknowledges support from FAPERJ (Grant No. 203.166/2017). I.S.O acknowledges FAPERJ (Grant No. 202.518/2019).

\section*{Acknowledgments}
We are in debt with Prof. Daniel Lidar and the Information Sciences Institute of the University of Southern California, for giving us access to the D-Wave Quantum Annealer that was used  in the first version of this work. The authors also acknowledges L. Cirto, N. L. Holanda and M.D. Correia for discussions and suggestions in the early stages of this work.

\section*{Supplemental Data}
 
\section*{Data Availability Statement}

\bibliographystyle{frontiersinHLTH&FPHY}
\bibliography{quantum_seismic}

\section*{Figure captions}

\begin{algorithm}
\begin{algorithmic}[1]
\Function{Solve}{$\mathbf{M}$, $\mathbf{t}$, $\mathbf{s_0}$, L, $N_{\mathrm{max}}$,$\epsilon$}  
   \State $\mathbf{I} \gets [1;1;1;\cdots;1]$ 
    \For {$c \gets 1$ to $N_{\mathrm{max}}$}
        \State  $\mathbf{b} \gets  (\mathbf{t}  + L \mathbf {M I} - \mathbf{M s_0 })/L$ 
        \State {\it Construct the vetor $\mathbf{q}$ and the Matrix $\mathbf{A}$}      
        \State {\it Convert $||\mathbf{A}\mathbf{q} - \mathbf{b}||^2 $ to QUBO  $\mathbf{q}^\mathsf{T} \mathbf{Q q}$} 
        \State {\it Calculate $\mathbf{q^*} $} 
        \State {\it From  $\mathbf{q^*}$  recover $\mathbf{x}$} 
        \State $\mathbf{s} \gets   \mathbf{s_0} + L\mathbf{(x-I)}$  
    
       \If {$||\mathbf{s} - \mathbf{s_{0}}||^2 \leq \epsilon$} 
            \State \emph{Stop Loop}  
        \Else  
           \State $\mathbf{s_0} \gets \mathbf{s}$ 
           \State $L \gets L/2$ 
        \EndIf 
    \EndFor 
 \State \emph{return} $\mathbf{s}$ 
\EndFunction 
\caption{ Function to solve the system of equations  $\mathbf{Ms} = \mathbf{t}$. The vector $\mathbf{s_0}$ is the initial guess for $\mathbf{s}$, $\epsilon$ is the tolerance,  $N_{max }$ is the maximum number of iterations, and $L$ defines the interval where we expect to find  the solutions (see text).  \label{recur}}
\end{algorithmic}

\end{algorithm}

\begin{figure}
\centering
 \includegraphics[width=1.0\linewidth]{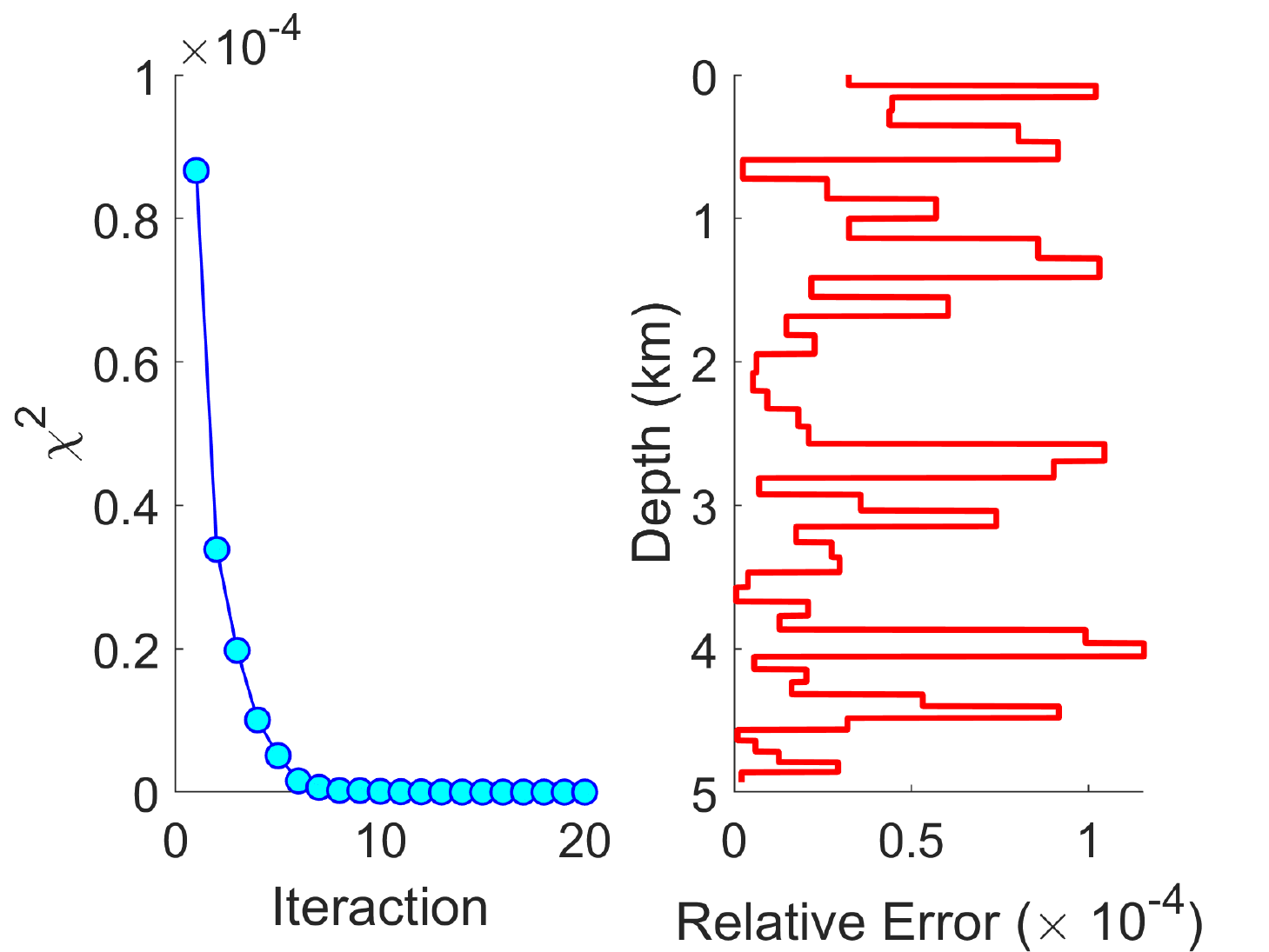}
\caption{Comparisons between classical and quantum solutions. In the left panel, we show the $\chi^2$ test between the quantum annealing solution and the classical solution. The relative error between the classical and quantum solutions is shown as a function of the ocean depth in the right panel. }
\label{conv}
\end{figure}

\end{document}